\documentclass[prl,twocolumn,showpacs
]{revtex4}
\newcommand{\be}{\begin{equation}}
\newcommand{\ee}{\end{equation}} 
\newcommand{\bea}{\begin{eqnarray}} 
\newcommand{\eea}{\end{eqnarray}}
\usepackage{graphicx}
\usepackage{epsfig}
\usepackage{bm}

\begin{document}
\title{Delta Hedged Option Valuation with Underlying Non-Gaussian Returns}
\author{L. Moriconi}
\affiliation{Instituto de F\'\i sica, Universidade Federal do Rio de Janeiro, \\
C.P. 68528, 21945-970, Rio de Janeiro, RJ, Brazil}
\begin{abstract}
The standard Black-Scholes theory of option pricing is extended to cope with 
underlying return fluctuations described by general probability distributions.
A Langevin process and its related Fokker-Planck equation are devised to model
the market stochastic dynamics, allowing us to write and formally solve the
generalized Black-Scholes equation implied by dynamical hedging. A systematic
expansion around a non-perturbative starting point is then implemented, recovering
the Matacz's conjectured option pricing expression. We perform an application of
our formalism to the real stock market and find clear evidence that
while past financial time series can be used to evaluate option prices before the
expiry date with reasonable accuracy, the stochastic character of volatility is an
essential ingredient that should necessarily be taken into account in analytical
option price modeling.
\end{abstract}

\maketitle

There has been a great interest in the study of the stochastic dynamics of financial markets
through ideas and techniques borrowed from the statistical physics context. A set of well-established 
phenomenological results, universally valid across global markets, yields the motivating ground for
the search of relevant models \cite{bouch-pott, mant-stan, voit}. 

A flurry of activity, in particular, has been related to the problem of option price valuation.
Options are contracts which assure to its owner the right to negotiate (i.e, to sell or to buy) for
an agreed value, an arbitrary financial asset (stocks of some company, for instance) at a future
date. The writer of the option contract, on the other hand, is assumed to comply with the option
owner's decision on the expiry date. Options are a crucial element in the modern markets, since
they can be used, as convincingly shown by Black and Scholes \cite{bs}, to reduce portfolio risk.

The Black-Scholes theory of option pricing is a closed analytical formulation where
risk vanishes via the procedure of dynamical (also called ``delta") hedging, applied to
what one might call ``log-normal efficient markets". Real markets, however, exhibit strong deviations
of log-normality in the statistical fluctuations of stock index returns, and are only approximately
efficient.

Our aim in this letter is to introduce a theoretical framework for option pricing which is general enough
to account for important features of real markets. We also perform empirical tests, taking the London market
as the arena where theory and facts can be compared. More concretely, we report in this work observational
results concerning options of european style, based on the FTSE 100 index, denoted from now on as
$S_t$ \cite{ft}. 

To start with, we note, in fact, that the returns of the FTSE 100 index do not follow log-normal
statistics for small time horizons. We have considered a financial time series of 242993 minutes 
(roughly, two years) ending on 17th november, 2005. The probability distribution function (pdf) $\rho(\omega)$ 
of the returns given by $\omega \equiv \ln(S_t/S_{t-1})$, taken at one minute intervals is shown in Fig.1. We
verify that the Student t-distribution with three degrees of freedom conjectured by Borland \cite{borland} provides
a good fit to the data. A slightly better fitting is produced if a smooth gaussian envelope is introduced to
truncate the distribution at the far tails, viz:
\be
\rho(\omega) \propto \frac{1}{(a^2+\omega^2)^2} \exp(-\frac{\omega^2}{2b^2}) \ . \ \label{r-pdf}
\ee

\begin{center}
\begin{figure}[tbph]
\hspace{-0.0cm} \includegraphics[width=9.0cm, height=7.5cm]{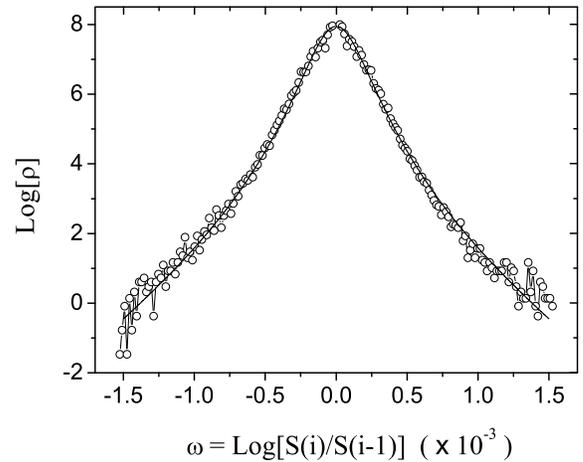}
\label{fig1}
\caption{The pdf of returns evaluated for one minute time horizons for the FTSE 100 index, along a period of about
two years, ending on 11/17/2005. The solid line is the truncated Student t-distribution given by Eq. (\ref{r-pdf}), with $a=2.3 \times 10^{-4}$ and $b/a=5$.}
\end{figure}
\end{center}

A time-dependent family of self-similar Tsallis distributions \cite{comment1}, $\rho (\omega,t) \propto (a^2(t)+\omega^2)^{1/(1-q)}$, with variance $a^2(t) \propto t^{2/(3-q)}$, was previously considered in option price modeling \cite{borland}. The time parameter $t$ refers here to the time horizon of return fluctuations, while $q \simeq 1.5$ is indicated from empirical pdfs. Even though suggestive results were found by Borland for a set of option prices taken from the real market, the time dependence of the variance $a^2(t)$ disagrees with observations. It is well-established that the variance of return fluctuations grows linearly with the time horizon within a certain range of days, and that the return pdfs are not self-similar (there is a crossover to the log-normal profile). A further difficulty has to do with the matching between the theoretical and observed volatilities. As reported in Ref. \cite{borland}, for instance, the model volatility needed to reproduce the volatility smiles of options based on the S$\&$P500 futures in june 2001 is set to $32.95 \%$. However, as the analysis of the S$\&$P500 futures time series reveals, the actual volatility for that period was considerably smaller, close to $14.5 \%$ \cite{comment2}. 

The essential reason for the specific choice of $a(t)$ in the above pdf is that it is related to a formerly known Langevin equation, which leads, on its turn, to a partial differential equation for option prices. An alternative and more general point of view -- to be pursued here -- is not to advance, {\it{a priori}}, any hypothesis on the form of the return pdfs, while still dealing with a Langevin description of return fluctuations.

Le us assume, therefore, that at time $t=0$ the underlying index is $S_0$ and that its subsequent values are given by
\be
S_t = S_0 \exp[ \mu t + x(t)]  \ , \ \label{lns-eq}
\ee
where $x(t)$ is a dynamic random variable described, at time $t$, by an arbitrary pdf $\rho(x,t)$. We take $x=0$ at $t=0$. Observe that it is a simple exercise to write, from Eq. (\ref{lns-eq}), the formal expression for the pdf of returns with time horizon $t$, in a statistically stationary regime.
 
As an inverse problem, we are now interested to find a function $f=f(x,t)$,  so that $\rho(x,t)$ be derived from the Langevin equation
\be 
\frac{d x}{dt} = f(x,t) \eta(t) \ , \ \label{lang-eq}
\ee
where $\eta(t)$ is a random gaussian field, defined by $\langle \eta(t) \rangle = 0$ and
$\langle \eta(t) \eta(t') \rangle = \delta(t-t')$. Actually, it is not difficult to compute $f(x,t)$
as a functional of $\rho(x,t)$. We just write down the Fokker-Planck equation that is satisfied by $\rho(x,t)$ \cite{borland2},
\be
\frac{\partial}{\partial t} \rho = \frac{1}{2} \frac{\partial^2}{\partial x^2} (f^2 \rho) \ , \ \label{fp-eq}
\ee
to obtain, from the direct integration of (\ref{fp-eq}),
\be
[f(x,t)]^2 = \frac{2}{\rho(x,t)} \int_{- \infty}^x dx' \int_{- \infty}^{x'} dx'' \frac{\partial}{\partial t} \rho(x'',t) \ . \ \label{f-def}
\ee
The function $f(x,t)$ is an important element in the option pricing problem, within the dynamical hedging strategy of investment.
Considering the writer of an option contract who owns a number $\Delta$ of the underlying asset, dynamical hedging, a concept firstly introduced by
Black and Scholes, consists of defining option prices $V = V(S,t)$ as a function of the asset's value $S$ and time $t$, so that $\Delta = \partial V / \partial S$, and the ``minimal" portfolio $\Pi(S,t) \equiv S \Delta(S,t) - V(S,t)$ is imposed to evolve according to the market risk-free interest rate $r$, as if $\Pi(S,t)$ were converted to cash and safely deposited into an interest earning bank account.
Under these conditions, a straightfoward application of Ito's lemma leads to the generalized Black-Scholes equation,
\be
\frac{\partial}{\partial t} V - r V + r S \frac{\partial}{\partial S} V + \frac{1}{2} f^2 S^2 \frac{\partial^2}{\partial S^2} V  = 0 \ . \ \label{gbs}
\ee
Exact solutions of this equation may be given in terms of statistical averages. For the case of call options which expire at time $t^*$, we get
\be
V =  \exp[-r(t^*-t)] \langle \Theta( S \exp( \zeta) - E) (S \exp (\zeta) - E ) \rangle  \ , \ \label{sol-gbs}
\ee
where $\zeta = \int_t^{t^*} dt (\dot x + r - \frac{1}{2} f^2) $ is a random variable described by some pdf $\phi=\phi(\zeta,T)$, with $T=t^*-t$ being the option's time to expire. One may check, by direct substitution, that (\ref{sol-gbs}) solves, in fact, Eq. (\ref{gbs}). The put-call parity relation is also satisfied by our solution, as any {\it{bona fide}} option pricing formula should do. It is worth noting that the case where $f(x,t)$ is constant corresponds to the standard Black-Scholes problem.

In order to develop analytical expressions for $V(S,t)$, let $Z(\lambda) = \langle \exp( i \lambda \zeta ) \rangle$ be the characteristic function associated to the distribution $\phi(\zeta,T)$. We may write, without loss of generality, $Z(\lambda) = Z_a(\lambda,0)Z_a(0,\lambda)Z_b(\lambda)$, where $Z_a(\lambda_1,\lambda_2) = \langle \exp[ i \lambda_1 (\int_t^{t^*} dt (\dot x + r) -\lambda_2 \frac{1}{2} \int_t^{t^*} dt f^2)] \rangle$. It is clear that $Z_a$ and $Z_b$ can be computed, in principle, with the help of the cumulant expansion. However, a non-perturbative analysis may be readily addressed when one realizes that $Z_a(\lambda,0)$ is nothing but the characteristic funtion of $\rho(x,T)$. The remaining problem, then, is to evaluate the corrections due to $Z_a(0,\lambda)$ and $Z_b(\lambda)$ (they are in fact small in realistic cases). Up to first order in $\lambda$, we have $Z_b(\lambda)=1$ and $Z_a(0,\lambda) = 1 - i \frac{\lambda}{2} \int_t^{t^*} dt \langle f^2 \rangle$. Using (\ref{f-def}), we get 
$Z_a(0,\lambda) = 1 - i \frac{\lambda}{2} \langle [x^2(t^*)-x^2(t)] \rangle \simeq \exp \{- i \frac{\lambda}{2} \langle [x^2(t^*)-x^2(t)] \rangle \}$. 
Taking now, as a phenomenological input, that the volatility depends linearly on the time horizon $T$, we write
$Z_a(0,\lambda) \simeq \exp [- i \frac{\lambda}{2} \sigma_0^2 T]$,
which implies that 
\be
\phi (\zeta, T) = \rho(\zeta - r T + \sigma_0^2 T /2 ,T) \ . \ \label{prob-dens1}
\ee
It is not difficult to verify that the above expression is exact when $f(x,t)=\sigma_0$, and leads to an option pricing
formula previously proposed by Matacz \cite{matacz}, which was based on heuristic arguments. However, the usefulness of the
approximation given by Eq. (\ref{prob-dens1}) is restricted to the cases where the far tails of $\rho(x,t)$ decay faster
than $|x|^{-1} \exp(-|x|)$, because of the exponential factor in Eq. (\ref{sol-gbs}). 

\begin{table}
\begin{center}
\begin{tabular}{|c|c|c||c|c||c|c|}
\hline
{Strike}
&\multicolumn{2}{c||}{02dec05}&\multicolumn{2}{c||}{06dec05}&\multicolumn{2}{c|}{09dec05} \\
\cline{2-7}
Price&MKT&EP&MKT&EP&MKT&EP\\
\hline\hline
$5125$  &  410.5 &  $ 412.67  $   & NA & X & NA & X \\
$5225$  & $ 312 $ & $ 312.79  $  & $324$ & $321.87 $ & $298$ & $297.51 $ \\
$5325$  & $ 214.5 $ & $213.94 $ & $ 225.5$ & $ 222.87 $ & $199$ & $197.72 $ \\
$5425$  & $ 122.5$ & $ 121.93  $  & $ 131.5$ & $ 129.48  $ & $103.5$ & $102.35 $ \\
$5525$  & 50 & $48.61$ &  53.5 &  $53.52 $ & $29.5$ & $29.72 $ \\
$5625$  & 13 & $13.01 $ &  12.5 &  $14.97 $ & $3.5$ & $4.79 $ \\
$5725$  & 2.5 & $[0.60]$ &  2 &  $1.66 $ &    $0.5$ & $0.35 $ \\
$5825$  & 0.5 & $[0.0]$ &  -- &  $0.0 $  &  -- & $0.0$ \\
$5925$  & NA & X &                 -- &  0.0   & -- & 0.0 \\
\hline
\end{tabular}
\end{center}
\caption{Listing of call option premiums taken on 02dec05 ( $S=5528.1$, $g=0.81$, $\sigma^*= 6.1 \%$), 06dec05 ($S=5538.8$, $g = 0.91$, $\sigma^* = 6.9 \%$), and 09dec05 ($S=5517.4$, $g = 0.94$, $\sigma^* = 7.1 \%$). The risk-free interest rate is $r=4.5\%$. 
Options expired on 16dec05. The mean volatility measured between 02dec05 and 09dec05 is $\sigma = 8.0 \%$.}
\end{table}

\begin{table}
\begin{center}
\begin{tabular}{|c|c|c||c|c||c|c|}
\hline
{Strike}
&\multicolumn{2}{c||}{19dec05}&\multicolumn{2}{c||}{03jan06}&\multicolumn{2}{c|}{12jan06} \\
\cline{2-7}
Price&MKT&EP&MKT&EP&MKT&EP\\
\hline\hline
$5225$  &  329.5 &  $ 331. 58  $   & NA & X & NA & X \\
$5325$  & $ 234.5$ & $ 235.87 $  & $368.5$ & $369.21 $ & $414$ & $415.84 $ \\
$5425$  & $ 148 $ & $148.07 $ & $ 271$ & $ 269.79 $ & $314$ & $315.92 $ \\
$5525$  & $ 76$ & $ 75.38  $  & $ 177$ & $ 176.42  $ & $215$ & $216.02 $ \\
$5625$  & 28.5 & $28.23$ &  93 &  $93.21 $ & $119$ & $119.24 $ \\
$5725$  & 8 & $[4.57]$ &  34.5 &  $36.36 $ & $40$ & $39.26 $ \\
$5825$  & 2.5 & $[0.43]$ &  9 &  $9.20 $ &    $5.5$ & $6.62 $ \\
$5925$  & 0.5 & $[0.0]$ &  2 &  $[0.52] $  &  0.5 & $0.34 $ \\
$6025$  & NA & X &                 0.5 &  [0.0]   & -- & 0.0 \\
\hline
\end{tabular}
\end{center}
\caption{Listing of call option premiums taken on 19dec05 ($S=5539.8$, $g=0.78$, $\sigma^* = 5.9 \%$), 03jan06 ($S=5681.5$, $g = 0.83$, $\sigma^*= 6.3 \%$), and 12jan06 ($S=5735.1$, $g = 0.78$, $\sigma^* = 5.9 \%$). The risk-free interest rate is $r=4.5\%$. Options expired on 20jan06. The mean volatility measured between 19dec05 and 12jan06 is $\sigma = 6.1 \%$.}
\end{table}

In a more empirically oriented approach, we may attempt to compute $V(S,t)$, as defined by Eq. (\ref{sol-gbs}), from real
or Monte Carlo simulated financial time series, without having any information on the pdf $\rho(x,t)$.
Since we would not be entitled to use Eq. (\ref{f-def}) anymore, it is necessary to rewrite the expression
$\int_t^{t^*} dt f^2$ that appears in the definition of $\zeta$ in terms of known quantities. Observing that Ito's lemma yields,
from Eq. (\ref{lang-eq}),
\be
e^{-x} \frac{d}{d t} e^{x} = \dot x + \frac{1}{2} f^2 \ , \
\ee
we write, for a time series of temporal length $T=N \epsilon$,
\be
\frac{1}{2} \int_t^{t^*} dt f^2 \simeq -(x_N -x_0)- [N- \sum_{i=0}^{N-1} \exp(x_{i+1}-x_i)] \ , \ \label{int-f}
\ee
where $x_i \equiv x(i \epsilon+t)$. Substituting (\ref{int-f}) into (\ref{sol-gbs}), we find, then, a pragmatical formula to work
with numerical samples.

\begin{center}
\begin{figure}[tbph]
\hspace{0.0cm} \includegraphics[width=9.7cm, height=14.0cm]{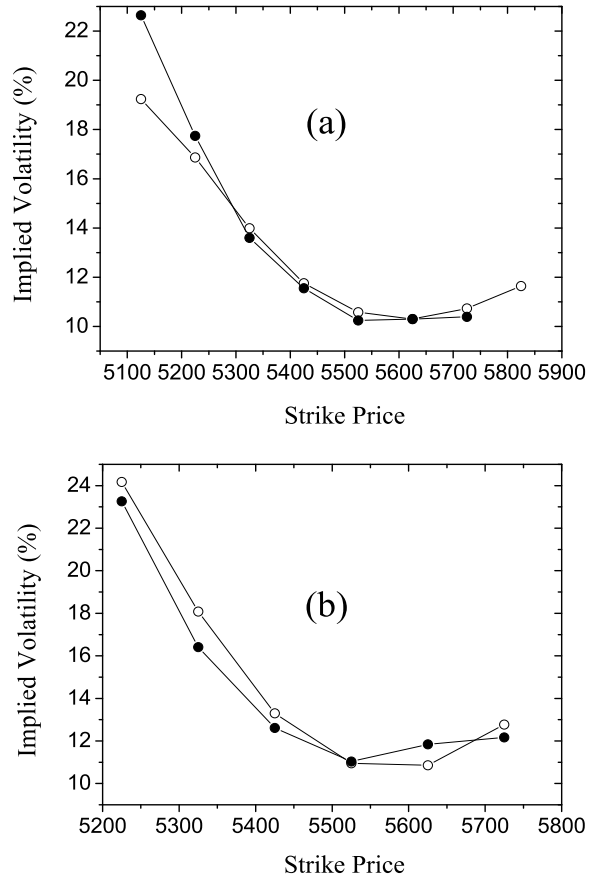}
\label{fig2}
\caption{Volatility smiles related to the option values taken on (a) 02dec05 and (b) 09dec05 (see Table I). The black and white dots refer to EP and MKT
premiums, respectively.}
\end{figure}
\end{center}

A stochastic process $\{x_i\}$ was generated from a high-frequency time series of the logarithms of FTSE 100 index, $\ln(S_i)$, consisting of 242990 minutes ending on 17th november, 2005 (it is essentially the same data used to establish the histogram depicted in Fig.1). The series was partitioned into 94 pieces of 2585 minutes each (roughly one financial week). In each one of these pieces, the market drift parameter $\mu$ was determined by the least squares method. Then, according to Eq. (\ref{lns-eq}), we define $\delta x_i \equiv x_{i+1}-x_i = \ln(S_{i+1}/S_i) - \mu$. We have found that improved results are obtained if extremely intense fluctuations are removed from the series, since they are likely to correspond to the market's reaction to unexpected events (we have checked this assumption in a number of cases). Once the standard deviation for the whole series is $\sigma_0 = 2.6 \times 10^{-4}$, we removed from the series the fluctuations given by $|\delta x_i| > 3 \times 10^{-3} \sim 10 \sigma_0$ (only 144 elements were taken out from the $\delta x_i$ series). A sequence of 3000 samples, separated in time by one hour translations, was used in each option price evaluation. It turns out that the samples are related to a period of reasonably well-behaved volatility (it fluctuates around some stable value). We also introduce a phenomenological factor $g$, so that trials with series of different mean volatilities are obtained in a simple way from the mapping $\delta x_i \rightarrow g \delta x_i$.

We have computed in this way option premiums based on the London FTSE 100 index (with $\epsilon =$ 1
minute). The results are shown in Tables I and II, where market (MKT) and evaluated prices (EP) are compared. In the tables,
NA stands for ``not available" data, while X stands for prices we did not evaluate; the values between brackets
correspond to instances where the time series extension was not large enough to get accurate predictions.
The daily closing values of the FTSE 100 index, $S$, the $g$-factors, and the model and observed mean volatilities ($\sigma^*$ and $\sigma$, respectively) are reported in the table captions. The numerical errors in the EP columns vary typically from $0.5 \%$ for the larger premiums to $10 \%$ for the smaller ones. When comparing the market and numerical results, one should keep in mind the existence of the usual bid-ask spread of option prices.

\begin{table}
\begin{center}
\begin{tabular}{|c|c|c||c|c||c|c|}
\hline
{Strike}
&\multicolumn{2}{c||}{02dec05}&\multicolumn{2}{c||}{06dec05}&\multicolumn{2}{c|}{09dec05} \\
\cline{2-7}
Price&BS&ST&BS&ST&BS&ST\\
\hline\hline
$5125$  &  409.38 &  $ 403.02  $   & X & X & X & X \\
$5225$  & $ 309.51$ & $ 303.15 $  & $318.91$ & $313.25 $ & $295.58$ & $291.10 $ \\
$5325$  & $ 209.64 $ & $203.30 $ & $ 219.02$ & $ 213.34 $ & $195.64$ & $192.17 $ \\
$5425$  & $ 110.92$ & $ 107.92  $  & $ 119.94$ & $ 116.60  $ & $96.30$ & $93.98 $ \\
$5525$  & 31.15 & $31.45$ &  36.76&  $37.19 $ & $18.44$ & $18.92 $ \\
$5625$  & 2.55 & $2.27$ &  3.51 &  $3.07 $ & $0.43$ & $0.24 $ \\
$5725$  & 0.04 & $0.02$ & 0.07 &  $0.07 $ &    $0.0$ & $0.0 $ \\
$5825$  & 0.0 & $0.0$ &  0.0 &  $0.0 $  &  0.0 & $0.0 $ \\
$5925$  & X & X &                0.0 &  0.0   & 0.0 & 0.0 \\
\hline
\end{tabular}
\end{center}
\caption{Comparison between Black-Scholes (BS) and Monte-Carlo simulations based on the Student t-distribution (ST).
We use the modeling parameters $(S,\sigma^*,r)$ reported in Table I. 
Expiry date is 16dec05 as well. }
\end{table}

As clearly shown in Fig. 2, we have been able to model volatility smiles which are in good agreement with the market ones. The remarkable feature of these results is that implied volatilities can be obtained from a single model volatility, $\sigma^*$, which is in fact close to the actual observed market volatility $\sigma$. 

We have, along similar lines, evaluated option prices through Monte Carlo simulations. The stochastic process $\{\delta x_i\}$ is now generated as independent events from the Student t-distribution with three-degrees of freedom. As shown in Table III, the Black-Scholes (BS) and Student t-distribution-based evaluations (ST) yield relatively close (and not good) answers for option prices. It is important to note the numerical results do not differ too much if Matacz's analytical option price formula is used with convolutions of the truncated Student t-distribution, Eq. (\ref{r-pdf}), once the log-normal return pdf cores get large enough for time horizons of a few days.

To summarize, we have provided strong evidence, from the empirical analysis of non-gaussian market data, that the combination of delta-hedging strategy and suitable Langevin modeling allows one to compute option premiums with reasonable confidence, from the use of Eqs. (\ref{sol-gbs}) and (\ref{int-f}). On the other hand, we have found that both the Monte Carlo simulations and the delta-hedging analytical framework based on fat-tailed distributions and time-independent volatilities which are close to the observed averaged values, fail to predict real market option prices. Our results point out that efficient option pricing analytical tools have necessarily to deal with the stochastic nature of volatility fluctuations, a main distinctive aspect of financial time series.

The author thanks Marco Moriconi for a critical reading of the manuscript. 
This work has been partially supported by FAPERJ.


\begin{references}

\bibitem{bouch-pott} J.-P. Bouchaud and M. Potters, {\it{Theory of Financial Risks - From Statistical Physics to Risk Management}},
Cambridge University Press, Cambridge (2000).

\bibitem{mant-stan} R. Mantegna and H.E. Stanley, {\it{An Introduction to Econophysics}}, Cambridge University Press, Cambridge (2000).

\bibitem{voit} J. Voit, {\it{The Statistical Mechanics of Financial Markets}}, Springer-Verlag (2003).

\bibitem{bs} F. Black and M. Scholes, J. Polit. Econ. {\bf{81}}, 637 (1973).

\bibitem{ft} Option premiums based on FTSE 100 index are daily published by the Financial Times at www.ft.com.

\bibitem{borland} L. Borland, Phys. Rev. Lett. {\bf{89}}, 098701 (2002).

\bibitem{comment1} A distribution $\rho(x,t)$ is called self-similar if there is, for any $t_0$, a time-dependent function $g=g(t)$
so that $\rho(x,t) = g(t) \rho(x g(t),t_0)$.

\bibitem{comment2} We refer, throughout the paper, to annualized volatilities, computed from the standard deviation of
one-minute returns, $\sigma_0$, as $\sigma = \sigma_0  \sqrt{8.5 \times 60 \times 252}$, where we assume 252 trading days
per year, and 8.5 market hours per day.

\bibitem{borland2} L. Borland, Phys. Rev. E {\bf{57}}, 6634 (1998).

\bibitem{matacz} A. Matacz, Int. J. Theor. Appl. Finance {\bf{3}}, 143 (2000).

\end{references}
\end{document}